\documentclass[twocolumn,showpacs,preprintnumbers,amsmath,amssymb]{revtex4}

\usepackage{graphicx}
\usepackage{dcolumn}
\usepackage{bm}
\begin{document}  
\title{Excited nucleons with  chirally improved fermions }
\author{Dirk Br\"ommel$^a$, Peter Crompton$^{b,a}$, 
Christof Gattringer$^{a}$, \\
Leonid Ya.\ Glozman$^c$, 
C.\ B.\ Lang$^c$, Stefan Schaefer$^a$ \\
and Andreas Sch\"afer$^a$\\
(for the BGR [Bern-Graz-Regensburg] Collaboration)}
\vskip1mm
\affiliation{$^a$ Institut f\"ur Theoretische Physik, Universit\"at
Regensburg,
D-93040 Regensburg, Germany}
\affiliation{$^b$ Departament de Fisica Teorica, 
Universitat de Valencia,
E-46100, Burjassot, Valencia, Espanya}
\affiliation{$^c$ Institut f\"ur Theoretische Physik, Universit\"at
Graz,
A-8010 Graz, Austria}
\begin{abstract}  
We study positive and negative parity nucleons on the  lattice using the
chirally improved lattice Dirac operator.  Our analysis is based on a set of
three operators $\chi_i$ with the nucleon quantum numbers but in different
representations of the  chiral group and with different diquark content. We use
a variational method to  separate ground state and excited states and determine
the mixing coefficients for the optimal nucleon operators in terms of the
$\chi_i$. We clearly identify the negative parity resonances $N(1535)$ and
$N(1650)$ and their masses agree well with experimental data. The mass  of the
observed excited positive parity state  is too high to be interpreted as the
Roper state.  Our results for the mixing coefficients  indicate that chiral
symmetry is important for $N(1535)$ and $N(1650)$ states.  We confront our data
for the mixing coefficients with  quark  models and provide insights into the
physics of the nucleon system  and the nature of strong decays.
\end{abstract}
\maketitle

\section{Introduction}\label{SectIntroduction}

Lattice QCD has the potential to answer some of the most important questions
about  the proper effective degrees of freedom in the nucleon and its
excitations.  In particular, one is able to arbitrarily change the mass of
quarks and hence to study the evolution of nucleon properties such as its mass
and excitation spectrum as a function of the current quark mass. This is
interesting  because we know the physical picture which drives hadrons made of
heavy quarks. In this case the hadrons can be  well approximated by the
non-relativistic QCD picture, where the most important ingredients are the
linear + Coulomb potential as well as a hyperfine correction that originates 
from the perturbative color-magnetic interaction between spins of heavy
quarks. 

The most prominent implication of this picture is the structure of the heavy
quarkonium spectrum. The color-magnetic spin-spin force induces a splitting
between the $^3S_1$ and $^1S_0$ ground states. The linear + Coulomb interaction
drives  the pattern of radially and orbitally excited states: The first orbital
excitation which carries $L=1$ orbital angular momentum and has a parity which
is opposite to the given ground state, lies always below the first radial
excitation, which has the same quantum numbers as the ground  state. Such a
pattern  is quantitatively understood within non-relativistic lattice QCD  as
well as the potential approach (for a review and references see e.g. Ref.
\cite{D}). 

On the other hand the ordering of the lowest positive and negative parity
excitations of the nucleon is just opposite. The first radial excitation of the
nucleon, the Roper resonance $1/2^+$, $N(1440)$, lies about 100 MeV below the
first orbital excitation, $1/2^-$, $N(1535)$. A similar pattern is observed in
the $\Delta$ spectrum, while the opposite ordering of the lowest 
positive-negative parity $J=1/2$ states takes place in the $\Lambda$ spectrum.
It appears to be impossible to explain this within a linear confinement plus
perturbative gluon exchange correction picture.

The whole experience of nuclear physics as well as numerous data on the chiral
structure of the nucleon teach us that the consequences of spontaneous chiral
symmetry breaking should be very important for the nucleon. A possible --
certainly oversimplified -- picture is that  {\it in the low energy regime} the
nucleon can be approximated as a system of three quasi-particles with dynamical
mass induced by spontaneous chiral symmetry breaking (constituent quarks) which
are confined. Such a picture is the most straightforward realization of the
$SU(6)$ symmetry. The latter, in turn, can be proved to be an exact symmetry of
QCD for ground state baryons in the large $N_c$ limit \cite{largeN}. Then the
question arises, what are the other residual interactions between the
constituent quarks which break $SU(6)$ symmetry?  The empirical ordering of
positive and negative parity states is opposite in mesons and baryons which
indicates that some physical aspects are different in these systems. An
anomalous ordering in baryons was explained to be possibly due to a significant
contribution of the residual interaction mediated by the chiral (Goldstone
boson) field \cite{G}. A meson-like exchange interaction between valence quarks
is possible within the quenched approximation in baryons, and not possible in
mesons.

Attempts to address these issues within the lattice formulation have begun only
recently 
\cite{manycalcs,blumetal,parityminussplita,parityminussplitb,leeetal,dongetal,sasaki}. 
Mainly the two interpolating fields 
$\chi_1$ and $\chi_2$  (for their definition see below) have been used in order
to extract the ground state nucleon as well as the lowest excited states of
both parities. The operator $\chi_1$ is known to couple well to the nucleon
ground state and has been used in lattice calculations for more than 20 years.
This operator also couples to the $J^P=1/2^-$ (negative parity) state. And
indeed, the lowest $1/2^-$ state was established and has a tendency to approach
the physical $N(1535)$ state for small quark mass. The authors of
\cite{blumetal} see no splitting between the lowest and the  next excited
negative parity state $J^P=1/2^-$ $N(1650)$, whereas such splitting is noticed
in  \cite{parityminussplita,parityminussplitb}. 

The first excited positive parity $N(1440)$ Roper state has not been identified
in lattice studies
\cite{manycalcs,blumetal,parityminussplita,parityminussplitb}. Initial hopes
that it would have a reasonable overlap with $\chi_2$ have been abandoned since
the plateaus seen from  the $\chi_2$ correlator come out consistently too high
(close to 2 GeV) to be the Roper state. 

Recently \cite{leeetal,dongetal} the Roper state has been observed and
reordering of the levels, such that the excited positive parity state falls
below the negative parity states taking place  at pseudoscalar masses somewhere
between 200 - 500 MeV, was obtained in 
\cite{leeetal,dongetal,sasaki} using the
$\chi_1$ interpolator. Both articles rely crucially on advanced fitting
techniques, i.e.\ Bayesian priors.

Here we tackle the analysis of the nucleon system   combining different
powerful strategies. In addition to the operators $\chi_1, \chi_2$ we use a
third operator $\chi_3$ and compute all cross correlations. This correlation
matrix is then analyzed using the  variational method \cite{variation} which
allows for a  separation  of ground and excited states. Using more operators in
the variational method  allows the system to better select those combinations
of interpolators which have optimal overlap with the physical  states. The
possibility of a reordering of the levels at small quark masses suggests to use
a Dirac operator with good chirality which allows one to go to small quark
masses. Here we work with the chirally improved Dirac operator \cite{chirimp}.
We perform a careful analysis of finite volume effects which is crucial in a
study of excited nucleons with their extended wavefunctions. 

From the variational technique one obtains also important information on the
optimal operators which have maximal overlap with the physical states. This is
interesting not only from a technical  point of view but it also provides
important insight into the  physics of the nucleon system. We compute the
mixing coefficients which relate our operators $\chi_i$ to the optimal
operators, i.e.\ the coefficients which determine the operator content of the
physical states.  These mixing coefficients are then studied as a function of
the quark mass and  confronted with information from quark models. We discuss
the  physical implications of this comparison. 

Our article is organized as follows: In the next section we collect
information on the lattice calculation such as simulation details, the
operators used, the variational technique and our fitting methods. In Section 3
we discuss the diquark content of our nucleon interpolators and their chiral
properties. This is followed by two sections where we discuss our raw data and 
then present the results for the mass spectrum.  In Section \ref{SectPhysCont} we
determine the mixing coefficients, discuss their physical implications in
Section \ref{SectMixingCoeffs}  and we end with conclusions in Section
\ref{SectConcl}.

\section{Lattice technology}\label{SectLattTech}

For our quenched calculation we use uncorrelated configurations from  the
L\"uscher-Weisz gauge action \cite{Luweact}. We work on $16^3 \times 32$ and
$12^3 \times 24$ lattices using 100 configurations for both volumes and
gauge coupling constant $\beta_{LW}=7.9$. The
lattice spacing is $a = 0.15$\ fm as determined from the Sommer parameter in
\cite{scale}. We remark that the scale from the Sommer parameter agrees well
with a determination  from hadronic input \cite{bgrlarge}. This gives rise to
spatial  extensions of 2.4 fm, respectively 1.8 fm. Throughout this paper we 
use this scale where we give mass values in physical units.

The gauge field configurations were treated with one step of hypercubic
blocking \cite{hypblocking}. We use the recently developed Chirally Improved
(CI) Dirac operator \cite{chirimp}. The CI operator is a systematic expansion
of a solution  of the Ginsparg-Wilson relation \cite{GiWi82} and has very good
chiral properties. In \cite{bgrlarge} the residual quark mass for the 
$\beta_{LW}= 7.9$ ensembles used here was computed from the chiral extrapolation of the
pion mass and found to be $m_{res} = -0.0020(5) a^{-1} = - 2.7(7)$ MeV. 
It was demonstrated \cite{bgrlarge,boston} that the  CI
operator allows simulations with pseudoscalar-mass to vector-mass  ratios down
to $m_{PS}/m_V$ = 0.28 at relatively small cost and that it has good scaling
properties. It was found \cite{bgrlarge} that for the nucleon the scaling
dependence on $a$ is of the size of the statistical error. For  the states we
focus on here, i.e. the excited positive parity state and the negative parity
states, we  have larger errors and the effects of finite $a$ are negligible. We
use Jacobi smeared sources \cite{jacobi} with a width of about 0.7 fm. For the
nucleon we find that the signal  at small quark masses is enhanced when
smearing also the sink. This allows to go deeper to the chiral limit, while at
the larger quark masses we  find exact agreement between the data from smeared
and point-like sinks.  Thus for the nucleon we use smeared sinks for all quark
masses.  All correlators were projected to zero momentum.   The inversion of
the Dirac operator was done with the BiCGstab multi-mass solver \cite{beat} and
we use several different values of the  quark mass in the range between $a\,m_q
= 0.013$ ($m_{PS}=270$ MeV) and  $a\,m_q = 0.2$ ($m_{PS}=866$ MeV). A more
detailed account of our setting is given in \cite{bgrlarge}. 

For the extraction of nucleon data we implement the following set of 
operators:
\begin{eqnarray}
\chi_1(x) & = & 
\; \;  \epsilon_{abc} \left[u_a^T(x)\,C \,\gamma_5\,d_b(x)\right]\,u_c(x) \; ,
\label{F1}
\\
\chi_2(x) & = &
\; \;  \epsilon_{abc} \left[u_a^T(x)\,C\, d_b(x)\right] \,\gamma_5 \,u_c(x) \; ,
\label{F2}
\\
\chi_3(x) & = & 
i \, \epsilon_{abc} \left[u_a^T(x)\,C\, 
\gamma_{\mu}\gamma_5d_b(x)\right]u_c(x) \; .
\label{F3}
\end{eqnarray}
$C$ is the charge conjugation matrix and $T$ denotes transposition. The color
indices are $a,b,c$, while for the Dirac indices we use matrix/vector notation.
For the interpolator $\chi_3$ we use only the time-like component.
All three operators have the quantum numbers  of the nucleon but correspond to 
different representations  of the chiral group and contain different diquark
fields in the brackets in Eqs.\ (\ref{F1}) - (\ref{F3}). 
Therefore one would expect that they couple
differently to the given physical state. In the next section we will discuss in
detail  these representations, the quantum numbers of the diquark fields  and
their physical implications.

We use the variational technique \cite{variation}  and compute all cross
correlations of our operators $\chi_i$ listed in  Eq.\ (\ref{F1}) - (\ref{F3}):
\begin{equation}
M_{\;i,j}^\pm(t) = \Big\langle \; \mbox{Tr}\, \Big[ \,
P_\pm \chi_i(t) \ 
\chi_j(0)^\dagger \, \Big] \; \Big\rangle \; \; .
\label{matrixM}
\end{equation}
We explicitly project our correlators to definite parity by inserting the
projectors $P^\pm = \frac{1}{2} [ 1 \pm \gamma_4 ]$ and the trace in  Eq.\
(\ref{matrixM}) is taken in Dirac space. Note that 4 is the euclidean time
direction and $\gamma_4$ the  corresponding element of the Clifford algebra. In
the following we  often omit the superscript $\pm$ for brevity. In order to
extract the ground state and excited levels we first normalize the correlation
matrix $M$ and construct $D(t) = M^{-1/2} (t_0) M(t) M^{-1/2} (t_0)$, where the
position $t_0$ of the time slice chosen for the normalization can be used to
optimize the signal. For all results presented here we have  used $t_0 = 1$,
which in our conventions corresponds to propagation distance 0. We discuss this
choice in Sect.\ref{SectResults}.

Note that $M$ is a hermitian, positive  definite matrix such that its inverse
square root exists. We remark that this normalization is designed to reduce the
error but a separation of ground and excited states can be obtained also
without this normalization. Actually we find that in our case we could work
almost equally well with the  unnormalized matrix $M$. The second step is a
diagonalization of the hermitian matrix $D(t)$  giving rise to three real
eigenvalues $\lambda_i(t), i = 1,2,3$.  Applying the lemma proven in
\cite{variation} one finds
\begin{eqnarray}
\lambda_i(t)  &=& C_i \, \exp\left[-(t-t_0)\,E_i\right] \nonumber\\
&&\times \Big[
1 + {\cal O}\Big( \exp\left[-(t-t_0)\Delta E_i\right] \Big) \Big] \; ,
\label{lambdat}
\end{eqnarray}
where $\Delta E_i$ is the difference of $E_i$ to the first omitted energy level,
i.e.\ $\Delta E_i = |E_i - E_4|$. We remark that the form
quoted in Eq.\ (\ref{lambdat}) is the expression  for the positive parity
states when using the matrix $M^+$, projected to  positive parity. For fitting
the negative parity states   one has to time-reverse the correlators, i.e.
replace $t$ by $T - t$, where $T$ denotes the temporal extension of the
lattice. The eigenvalues from  $M^-$ are related to the eigenvalues from $M^+$
by a minus sign and time reversal. We average our results from $M^+$ and $M^-$
after  the necessary transformation to increase the statistics.  

Eq.\ (\ref{lambdat}) makes clear the strength of the variational method: Each
eigenvalue corresponds to a different energy level  which dominates its
exponential decay.  For a large enough set of operators this allows for a clean
separation  of ground state and excited levels and we can apply a simple stable
2-parameter ($C_i,\,E_i$) fit to determine the corresponding energy. In
principle an $N\times N$ correlation matrix is sufficient for determining the
ground state and the first $N-1$ excited states as long as the operators are
linearly independent and each of the physical states has overlap with at least
one operator. However, using more operators improves the results. The
``perfect operator'' has infinitely many different lattice operators (point
split, different levels of smearing  etc.). Some have large coefficients and
others don't. Selecting those with large overlap is a highly non-trivial task
and requires in principle a knowledge of the wave function.

For a somewhat simpler problem - the application of the correlation matrix
technique to a scattering problem - where there is a natural ordering of the
operators (relative momenta of the scattering particles) a criterion to
determine how many operators are needed was given in \cite{variationb}.

Our $\chi^2$ functional uses the full correlation matrix.   The fit ranges were
determined from effective mass plateaus and  we find values of $\chi^2/d.o.f.$
ranging from 0.5 to 1.5. Slight  increases or decreases of the fit range leave
our fit results and the $\chi^2/d.o.f.$ essentially invariant. We  found that
in the negative parity sector and for the excited  positive parity state the
quality of the data decreases for  small quark masses and we could not maintain
the standards of our  fitting procedure. We do not give our results for these
cases.    The errors we quote were computed using jackknife subensembles.

\section{Diquark content of nucleon interpolators and their 
chiral properties}\label{SectDiquark}

Let us now discuss in more detail the structure of our nucleon operators
$\chi_i$ defined in Eqs.\ (\ref{F1}) - (\ref{F3}). The fields $\chi_1$ and
$\chi_2$ create from the vacuum both, positive and negative parity states, with
$I,\,J = 1/2,\,1/2$. Linear combinations of these fields have intensely  been
used in QCD sum rule calculations \cite{IOFFE}. The field $\chi_1$ contains a
scalar diquark \footnote{Here and in what follows "diquark" refers to a
subsystem of two quarks (the structure in  brackets in Eqs.\ (\ref{F1}) -
(\ref{F3})) without necessarily implying a quark-diquark clustering in the
baryon.} $I,\,J^P = 0,\,0^+$ component, while the field $\chi_2$ has a
pseudoscalar diquark component $I,J^P = 0,\,0^-$.

It is instructive to review the chiral $SU(2)_L \times SU(2)_R$  properties of
these fields.  Here and in the following $(I_L,\,I_R)$ denotes the irreducible
representations of $SU(2)_L \times SU(2)_R$ with $I_L$ and $I_R$ being isospins
of the left-handed and right-handed components. The quark field $q(x)$ (with $q
= u$ or $d$) together with its parity partner $\gamma_5 q(x)$ forms a
$(0,\,1/2) \oplus (1/2,0)$ reducible representation of the chiral group. This
reducible representation is an irreducible representation of the full
parity-chiral group $SU(2)_L \times SU(2)_R \times C_i$, where the group $C_i$
consists of the elements identity and space inversion \cite{CG}.

Consider first the scalar diquark field in (\ref{F1}). This diquark field is
invariant with respect to both isospin as well as axial transformations  in the
isospin space and thus transforms as a scalar representation $(0,\,0)$ of the
chiral group. The chiral transformation properties of the interpolator $\chi_1$
are then determined exclusively by the  transformation properties of the third
quark. Hence the interpolator $\chi_1$ together with its parity partner 
$\gamma_5 \,\chi_1$ transform as  $(0,\,1/2) \oplus (1/2,\,0)$,
\begin{equation}
\left\{\chi_1(x),\, \gamma_5 \,\chi_1(x)\right\} \sim (0,\,1/2) \oplus
(1/2,\,0) \; .
\label{G1}
\end{equation}
Similarly, the pseudoscalar diquark in $\chi_2$ also is invariant with respect
to chiral transformations. Again the transformation properties of $\chi_2$ are
solely determined by the transformation properties of the third quark field, 
\begin{equation}
\left\{\chi_2(x), \gamma_5 \chi_2(x)\right\} \sim (0,\,1/2) \oplus (1/2,\,0) \; .
\label{G2}
\end{equation}
We stress that the fields  $\{\chi_1(x),$ $\gamma_5\, \chi_1(x)\}$
and  $\{\chi_2(x),$ $\gamma_5 \,\chi_2(x)\}$ belong to {\it
distinct} {\it irreducible} representations of the parity-chiral group and they
do not transform into each other under chiral transformations.  Linear
combinations of the fields $\chi_1$ and $\chi_2$, as used in the QCD sum rule
approach,  form only {\it reducible} representations of the parity-chiral
group \footnote{The interpolators considered in Ref. \cite{CJ} correspond to
$\chi_1 \pm \chi_2$.}.

The transformation properties with respect to a flavor-singlet $U(1)_A$
transformation are quite different, because the flavor-singlet axial
transformation mixes scalar and pseudoscalar diquarks. Only the specific linear
combinations
\begin{eqnarray}
\Big\{ \left(\chi_1(x)+\chi_2(x)\right) &,&
i\gamma_5 \,\left(\chi_1(x)+ \chi_2(x)\right)\Big\} \; , 
\label{A1}
\\
\Big\{ \left(\chi_1(x)-\chi_2(x)\right) &,&
i\gamma_5 \,\left(\chi_1(x)- \chi_2(x)\right)\Big\} \; , 
\label{A2}
\end{eqnarray}
form distinct irreducible representations of $U(1)_A \times C_i$.

There is no reason to expect that the interpolators $\chi_1$ and $\chi_2$ are 
the only ones that couple to the nucleon and its excited states.  Hence {\it
a-priori} one should expect that the nucleon and its excited states are
actually some mixtures of different irreducible representations. In order to
check this aspect we also included our interpolator $\chi_3$.

In $\chi_3$ the diquark field in the brackets has  isoscalar-vector quantum
numbers $I,J^P = 0,\,1^-$ for the spatial components ($\mu=1,2,3$) of the
4-vector and isoscalar-scalar quantum numbers $I,J^P = 0,\,0^+$ for the time
component ($\mu=4$). Hence for $\mu=4$ this operator couples only to $J=1/2,
I=1/2$ states, similar to  $\chi_1$ and  $\chi_2$, whereas for $\mu=1,2,3$ it
couples to both $J=1/2$ and $J=3/2$  (of both isospins $I=1/2$ and $I=3/2$)
baryons. Contrary to the isoscalar-scalar diquark in $\chi_1$ the diquark field
in $\chi_3$ transforms as $(1/2,\,1/2)$ under $SU(2)_L\times SU(2)_R$ and these
transformation properties distinguish it from the diquark in $\chi_1$. This
also implies that $\chi_3$ together with some other fields generates the
$(1,\,1/2) \oplus (1/2,\,1)$ and $(0,\,1/2) \oplus (1/2,\,0)$ 
representations of the chiral group, the latter is
however distinct from the representations generated by $\chi_1$ and $\chi_2$.

\section{Discussion of the raw data}\label{SectRawdata}

Before presenting results for the masses let us briefly discuss our raw  data.
This discussion illuminates the roles of the different interpolators $\chi_i$
and underlines the strength of the diagonalization method.  We base this
analysis on effective masses computed from  either the independent diagonal
correlation functions  $\langle \chi_i \chi_i \rangle$ or from the eigenvalues
$\lambda_i$.

\begin{figure*}[t]
\includegraphics*[height=14.5cm]{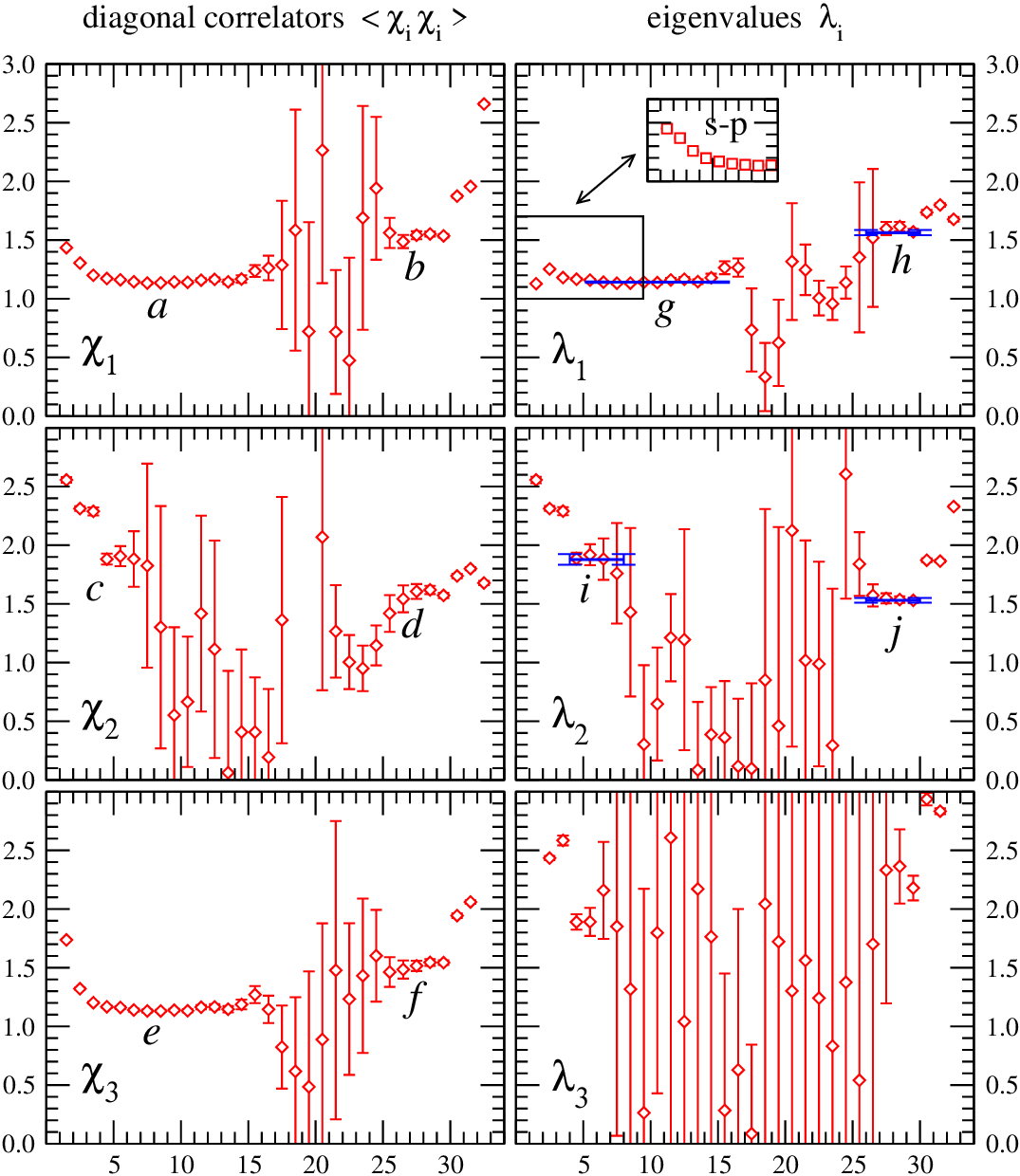}
\caption{Effective mass plots for the $16^3\times 32$ ensemble
at $a\,m_q=0.10$. The lhs.\ column shows results from the independent diagonal
correlators $\langle \chi_1 \chi_1 \rangle$, $\langle \chi_2 \chi_2 \rangle$
and $\langle \chi_3 \chi_3 \rangle$ (top to bottom) while the rhs.\ column is
for the eigenvalues $\lambda_i$ of the correlation matrix. The horizontal bars 
show our fitting results and range. The characters in the plots label
structures which we discuss in the text. The insert in the top right panel
demonstrates the effect of turning off the smearing of the sink.
\label{fig1}}
\end{figure*}

In Fig.\ \ref{fig1} we show in the left-hand-side (lhs.)~column   plots of
effective masses as  obtained from the independent diagonal correlators 
$\langle \chi_1 \chi_1 \rangle$, $\langle \chi_2 \chi_2 \rangle$ and $\langle
\chi_3 \chi_3 \rangle$ (top to bottom). The right-hand-side (rhs.)~column 
shows effective masses for the three  eigenvalues $\lambda_1$, $\lambda_2$ and
$\lambda_3$ of the correlation matrix. The horizontal bars shown  in the plots
for $\lambda_1$ and $\lambda_2$ indicate our fit results and the fitting range.
The characters in the plots  label structures which we discuss below. All data
are from  the $16^3 \times 32$ lattice at $a\,m_q = 0.10$. 

Let us first discuss the independent diagonal correlators  $\langle \chi_i
\chi_i \rangle$ shown in the lhs.\ column.  The positive parity states are
extracted from the left part ($t < 18$)  of the plots, while the negative
parity states that couple to the same interpolator dominate the right part ($t
> 18$) of the plots. One clearly sees a good plateau for the interpolators
$\chi_1$ and $\chi_3$ in the interval (5,\,15). This plateau corresponds to the
ground state nucleon (positions $a$ and $e$). At small Euclidean times one can
see contributions from the excited states of positive parity. If an
interpolator couples strongly to the nucleon ground state (as it is the case
for $\chi_1$ and $\chi_3$) then the signal from the excited states is only weak
and the extraction of the  mass for the excited states becomes a complicated
task.

The signal obtained with the interpolator $\chi_2$ is intriguing.  There is no
plateau corresponding to the nucleon ground state.  This fact has a natural
explanation within a quark model and will be discussed below. However, for
relatively large quark masses $a\, m_q \geq 0.08$ one sees a plateau at smaller
distances (3,\,7), i.e.\ position $c$. We remark that for the $a\, m_q = 0.10$ 
shown in the figure the quality of this plateau is already very poor,  but it
is more pronounced at our larger quark masses. The mass corresponding to this
plateau is higher than the mass of the negative parity state seen from the same
interpolator.  Though we do not know the fate of this excited positive parity
state at small  quark masses, it is unlikely to be the Roper (cf.\ our
discussion in Section  \ref{SectMixingCoeffs}).

Let us now look at the signals for negative parity states from the diagonal
correlators $\langle \chi_i \chi_i \rangle$. All three of them  show plateaus
for time intervals (25,30), i.e.\ positions $b, d, f$. Similar to the plateaus
for the excited positive parity state  these plateaus are more pronounced for
larger quark masses and vanish entirely below $a\,m_q = 0.06$. It is
interesting to note that the plateau in $\langle \chi_2 \chi_2 \rangle$
(position $d$) is somewhat higher  than the other two plateaus (positions $b,
f$). This is consistent with  an observation we present below where we will
show that at relatively large quark masses the  heavier negative parity state
$N(1650)$ has a larger $\chi_2$ component, while this component is smaller for
the lighter $N(1535)$.

The different states which appear in the individual diagonal correlators are
separated when analyzing the eigenvalues of the correlation matrix. The
corresponding effective mass plots are shown in the rhs.\ column of  Fig.\
\ref{fig1}. We order the eigenvalues $\lambda_i(t)$ with respect  to their
size, such that for each value of $t$ we have  $\lambda_1(t) > \lambda_2(t) >
\lambda_3(t)$. The nucleon ground state gives rise to a well pronounced plateau
in $\lambda_1$ in the time interval $(5,15)$ (position $g$). For the first
excited positive parity state there is a hint of a plateau in the range (4,8)
(position $i$),  but the quality is as poor as for the correlator  $\langle
\chi_2 \chi_2 \rangle$ alone. We remark that there is no sign of the excited
states contribution seen in the $\lambda_1$ plot at small times. The quality of
the plateaus  for the two negative parity states (positions $h$ and $j$) is
somewhat better. These plateaus become more pronounced for larger quark masses
but  vanish entirely for $a\,m_q < 0.06$. As is obvious from the bottom right 
plot the third eigenvalue $\lambda_3$ is highly contaminated with excited 
states, has large statistical errors, and no structure is visible in  the
corresponding effective mass plots.

For our variational analysis the number of independent operators should  be
sufficiently large and was limited to three for practical reasons. However,
we also check our results
by analyzing also the ($\chi_1$, $\chi_2$)-subsector. It
turns out that the resulting eigenvalues are close to those in  Fig.\
\ref{fig1}. Actually  $\chi_3$  seems to contribute to the same states as
represented by the other two operators  and no clear signal to further states
is observed. This is further discussed in Sect. \ref{SectMixingCoeffs}, but
also seen in Fig.\ \ref{fig1}, where $\lambda_3$ shows no additional plateau
signal.

\begin{figure}[t]
\includegraphics[width=6.3cm]{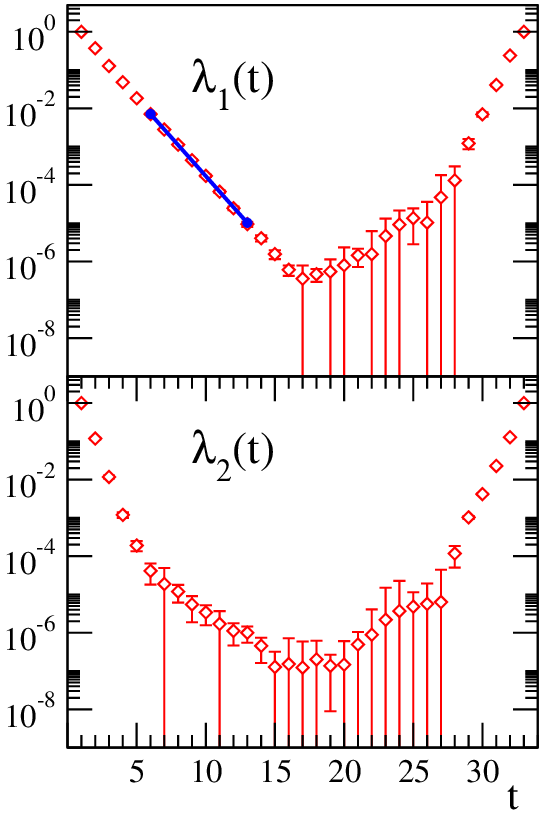}
\caption{The eigenvalues $\lambda_1(t)$ and  $\lambda_2(t)$ on a logarithmic
scale. The data are for $16^3 \times 32$ at $a\,m_q=0.04$). The full line in
the $\lambda_1$-plot (top) shows the fit for the nucleon mass.
\label{fig2}}
\end{figure}

In the insert (window) in the upper right plot of Fig.\ \ref{fig1} we show the
effect of  omitting the Jacobi smearing at the sink. All the correlators shown
in  Figs.\ \ref{fig1} and \ref{fig2} are obtained with both smeared source and
sink. The Jacobi smearing is designed to improve the overlap with the  ground
state and the insert demonstrates clearly, that the contribution of excited
states increases when omitting the smearing of the sink.  For the point-type
sink one now observes a stronger signal from excited states at small time. Note
also that since this signal appears in the $\lambda_1$ plot, and the
$\lambda_2$ plot still exhibits a plateau (which originates from the $\chi_2$
correlator), we  conclude that this signal comes from the $\chi_1$ interpolator
and the corresponding excited state is not the same as seen with the $\chi_2$
interpolator. We have tried to analyze the $\chi_1$ correlator with  two- and
multi-exponential fits in an attempt to extract this positive parity excited
state. However, the statistical errors are too large for a reliable extraction
and we cannot draw firm conclusions about the ordering relative to the negative
parity state.  Hence we cannot decide from such a fit about the energy of the
lowest positive parity excited state seen with the smeared source and point
sink. While we cannot conclude anything concrete whether this signal comes from
the Roper resonance or not, the insert shows  that optimizing the spatial
structure of the nonlocal correlators (in order to possibly separate the
signals from the nucleon and the Roper) and introducing more operators is one
of the possibilities for further lattice studies of the Roper state.

\begin{figure}[t]
\includegraphics[width=8.5cm,clip]{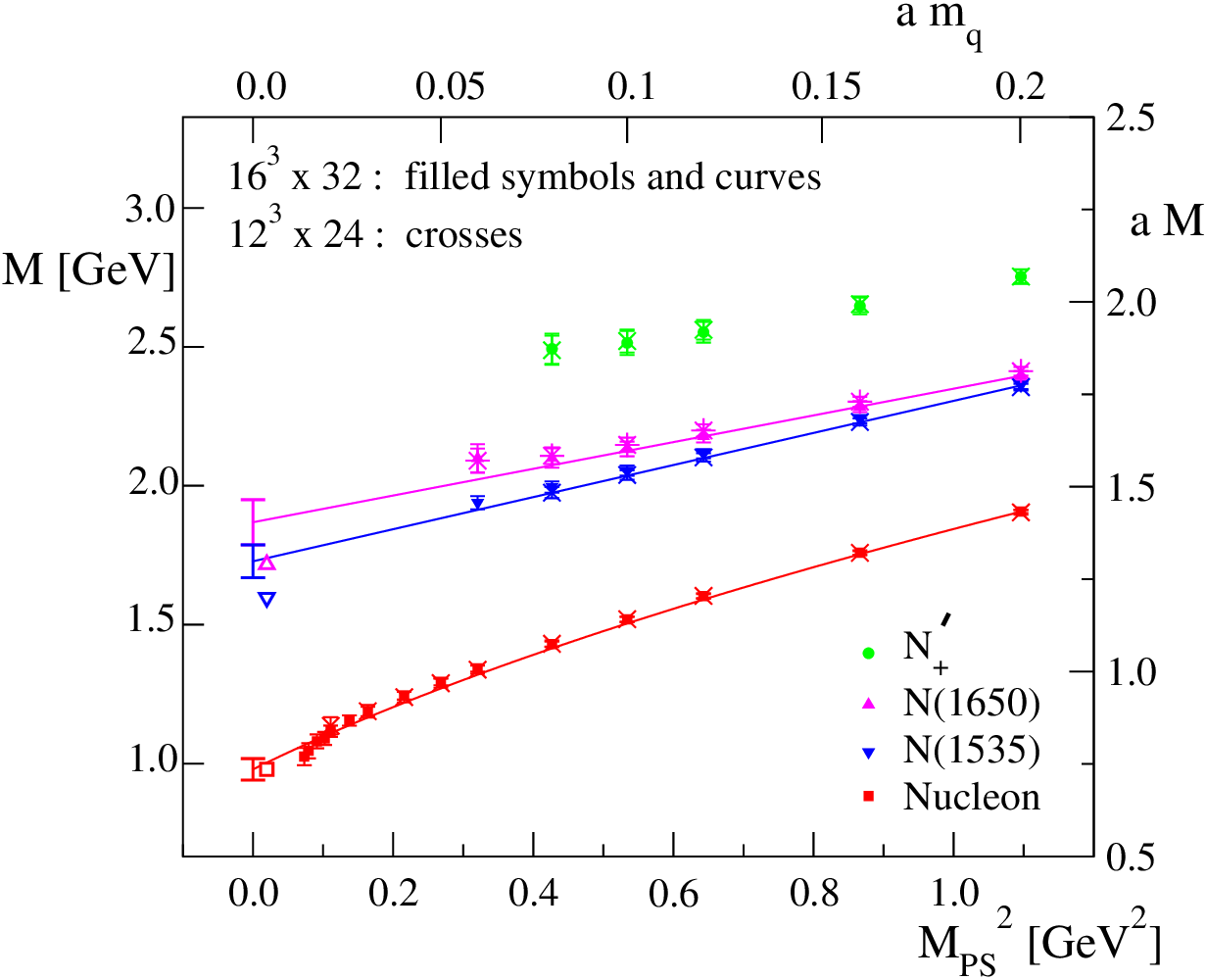}
\caption{ 
Masses of positive and negative parity states in GeV and in
lattice units as a function of
the pseudoscalar mass squared in $\textrm{GeV}^2$ (quark mass in lattice units
on top scale). We show data from  $16^3 \times 32$ lattices as filled symbols
and  compare them to data from $12^3 \times 24$ lattices (crosses). The 
curves represent the chiral extrapolation of the nucleon and
the two negative parity states. The open symbols show the experimental values.
\label{nucleonmasses}}
\end{figure}

Let us finally comment on an interesting structure which we observe  both in
the $\chi_2$ correlator and in $\lambda_2$ for small quark masses. In Fig.\
\ref{fig2} we illustrate this effect  by comparing $\lambda_1(t)$ and
$\lambda_2(t)$ on a logarithmic scale for  the $16^3 \times 32$ ensemble at
$a\, m_q = 0.04$. The top plot  shows $\lambda_1(t)$ and the full line
indicates where we fitted the nucleon mass. The behavior of the second
eigenvalue $\lambda_2(t)$ (bottom plot) shows a somewhat surprising behavior: It
starts off with a large slope but near $t=7$ turns into a nearly linear
behavior with a much smaller slope, in particular its absolute value is 
smaller than the one for the nucleon. If one interprets this as the trace of  a
physical positive parity particle it would be lighter than the nucleon.  As a
consequence we cannot  interpret this structure as a signal from a physical
state, but likely it is related to some quenched artifact. We remark that
also   the time behavior of the corresponding eigenvector  (see Sect.
\ref{SectPhysCont}) shows a change near $t=7$ indicating that the left half of
the $\lambda_2(t)$ does not  describe a proper physical state.

\section{Results for the mass spectrum}\label{SectResults}

\begin{table*}[Ht]
    \caption{Fit results of fits to $\lambda_1$ and $\lambda_2$. The nucleon
      ground state is obtained from smeared--smeared results, whereas the
      remaining states are from smeared--point calculations. The mass values
      are in lattice units, the fit interval
      is denoted by $t$.\label{table:parameters}}
\begin{ruledtabular}
\begin{tabular}[t]{lcccclccc}
      \multicolumn{4}{l}{$\mathbf{16^3 \times 32}$,
        $\mathbf{\boldsymbol{\beta_{LW}} = 7.90}$}&&
       \multicolumn{4}{l}{$\mathbf{12^3 \times 24}$,
        $\mathbf{\boldsymbol{\beta_{LW}} = 7.90}$}\\

      $a\,m_{\textrm{q}}$ & $t$ & $a\,M_{N}$ &
      $\chi^2_{\textrm{d.o.f.}}$\rule{0pt}{2.5ex}&&
      $a\,m_{\textrm{q}}$ & $t$ & $a\,M_{N}$ &
      $\chi^2_{\textrm{d.o.f.}}$\rule{0pt}{2.5ex}\\[.4ex]
      \hline
0.013 & $[5, 9]$ &  0.770 (23)  &  1.31  & &\multicolumn{4}{c}{\phantom{$[16,22]$}\rule{0pt}{2.3ex}}\\
0.014 & $[5,10]$ &  0.786 (20)  &  1.31  & &\multicolumn{4}{c}{\phantom{$[16,22]$}}\\
0.016 & $[5,11]$ &  0.811 (19)  &  1.47  & &\multicolumn{4}{c}{\phantom{$[16,22]$}}\\
0.018 & $[6,12]$ &  0.826 (16)  &  1.27  & &\multicolumn{4}{c}{\phantom{$[16,22]$}}\\
0.02  & $[6,12]$ &  0.839 (15)  &  1.28  & & 0.02  & $[5, 9]$ &  0.853 (23)  &  1.26 \\ 
0.025 & $[6,12]$ &  0.868 (14)  &  1.18  & &\multicolumn{4}{c}{\phantom{$[16,22]$}}\\
0.03  & $[6,13]$ &  0.892 (13)  &  1.39  & & 0.03  & $[5,10]$ &  0.893 (17)  &  0.58 \\ 
0.04  & $[6,13]$ &  0.935 (11)  &  1.01  & & 0.04  & $[5,11]$ &  0.931 (14)  &  0.25 \\ 
0.05  & $[6,14]$ &  0.972 (09)  &  0.75  & & 0.05  & $[5,12]$ &  0.969 (12)  &  0.43 \\ 
0.06  & $[6,14]$ &  1.008 (08)  &  0.66  & & 0.06  & $[5,12]$ &  1.005 (11)  &  0.53 \\ 
0.08  & $[6,14]$ &  1.077 (07)  &  0.73  & & 0.08  & $[5,12]$ &  1.075 (09)  &  0.75 \\ 
0.10  & $[6,15]$ &  1.141 (06)  &  1.07  & & 0.10  & $[5,12]$ &  1.142 (08)  &  0.93 \\ 
0.12  & $[6,15]$ &  1.204 (05)  &  1.12  & & 0.12  & $[5,12]$ &  1.205 (07)  &  1.05 \\ 
0.16  & $[6,15]$ &  1.321 (05)  &  1.15  & & 0.16  & $[5,13]$ &  1.321 (07)  &  1.22 \\ 
0.20  & $[6,16]$ &  1.433 (04)  &  1.34  & & 0.20  & $[5,13]$ &  1.431 (06)  &  1.09 \\[.3ex]
      \hline
      $am_{\textrm{q}}$ & $t$ & $aM_{N'_+}$ &
      $\chi^2_{\textrm{d.o.f.}}$\rule{0pt}{2.5ex}&&
      $am_{\textrm{q}}$ & $t$ & $aM_{N'_+}$ &
      $\chi^2_{\textrm{d.o.f.}}$\rule{0pt}{2.5ex}\\[.4ex]
      \hline
0.08  & $[4, 7]$ &  1.873  (41)  &  0.15 & & 0.08  & $[4, 8]$ &  1.869  (39)  &  0.04\rule{0pt}{2.3ex} \\
0.10  & $[4, 8]$ &  1.889  (32)  &  0.17 & & 0.10  & $[4, 8]$ &  1.894  (31)  &  0.96 \\ 
0.12  & $[4, 9]$ &  1.918  (28)  &  0.26 & & 0.12  & $[4, 8]$ &  1.924  (26)  &  0.94 \\ 
0.16  & $[4, 9]$ &  1.989  (22)  &  0.70 & & 0.16  & $[4, 8]$ &  1.993  (21)  &  1.12 \\ 
0.20  & $[4, 9]$ &  2.068  (19)  &  1.46 & & 0.20  & $[4, 8]$ &  2.069  (18)  &  1.66 \\[.3ex]
      \hline                                                            
      $am_{\textrm{q}}$ & $t$ & $aM_{N(1535)}$ &
      $\chi^2_{\textrm{d.o.f.}}$\rule{0pt}{2.5ex}&&
      $am_{\textrm{q}}$ & $t$ & $aM_{N(1535)}$ &
      $\chi^2_{\textrm{d.o.f.}}$\rule{0pt}{2.5ex}\\[.4ex]
      \hline
0.06  & $[27,30]$ &  1.457  (18)  &  0.04  & & \multicolumn{4}{c}{\phantom{$[16,22]$}}\\
0.08  & $[26,30]$ &  1.498  (15)  &  0.10  & &  0.08  & $[19,22]$ &  1.484  (15)  &  0.22\rule{0pt}{2.3ex} \\
0.10  & $[26,30]$ &  1.544  (14)  &  0.08  & &  0.10  & $[18,22]$ &  1.532  (13)  &  0.09 \\
0.12  & $[26,30]$ &  1.590  (12)  &  0.22  & &  0.12  & $[17,22]$ &  1.579  (12)  &  0.21 \\
0.16  & $[25,30]$ &  1.684  (11)  &  1.10  & &  0.16  & $[17,22]$ &  1.675  (10)  &  1.11 \\
0.20  & $[25,30]$ &  1.776  (10)  &  2.02  & &  0.20  & $[16,22]$ &  1.769  (09)  &  2.25 \\[.3ex]
      \hline
      $am_{\textrm{q}}$ & $t$ & $aM_{N(1650)}$ &
      $\chi^2_{\textrm{d.o.f.}}$\rule{0pt}{2.5ex}&&
      $am_{\textrm{q}}$ & $t$ & $aM_{N(1650)}$ &
      $\chi^2_{\textrm{d.o.f.}}$\rule{0pt}{2.5ex}\\[.4ex]
      \hline
0.06  & $[27,30]$ &  1.577  (38)  &  0.13 & &  0.06  & $[18,22]$ &  1.570  (33)  &  1.03\rule{0pt}{2.3ex} \\
0.08  & $[27,30]$ &  1.578  (26)  &  0.07 & &  0.08  & $[18,22]$ &  1.584  (23)  &  0.32 \\
0.10  & $[26,30]$ &  1.604  (21)  &  0.39 & &  0.10  & $[18,22]$ &  1.613  (19)  &  0.13 \\
0.12  & $[26,30]$ &  1.638  (18)  &  0.39 & &  0.12  & $[17,22]$ &  1.652  (16)  &  0.32 \\
0.16  & $[25,30]$ &  1.716  (15)  &  0.37 & &  0.16  & $[17,22]$ &  1.730  (14)  &  0.47 \\
0.20  & $[25,30]$ &  1.800  (13)  &  0.30 & &  0.20  & $[16,22]$ &  1.813  (12)  &  0.53 \\[.3ex]
    \end{tabular}
    \end{ruledtabular}
\end{table*}

Let us now come to the presentation of the nucleon masses which we extract from
our correlators. In Fig.\ \ref{nucleonmasses} we show the baryon masses in
lattice units as a function of the quark mass. Only data points were included
where the  standards of our fits discussed in Section \ref{SectLattTech} could
be maintained  and where a credible signal in the effective mass plot is seen.
They all come from the diagonalization approach using $t_0 = 1$.  We did try
$t_0$ up to 3. The quality of the plateaus does not improve and the results
remain invariant. Note that we use Jacobi smeared sources and there we find
that these effects are minimal. This behavior changes for more extended
sources. All fit results and parameters are listed in Table 
\ref{table:parameters}.

The lowest lying set of data  corresponds to the ground state in the positive
parity sector, i.e. the nucleon. This is followed by two negative parity states
which upon  simple chiral  extrapolation we identify as $N(1535)$ and
$N(1650)$. For brevity we use  this particle data book nomenclature in the
following. Although the two  negative parity states  are relatively close to
each other we  clearly  separate them with the variational technique discussed
above. The fact that the two  negative parity states become approximately
degenerate towards larger quark masses is well understood in potential models
for heavy quarks.  Finally, the excited state  with positive parity is found on
top and we will refer to it as $N_+^\prime$.  While we can identify the nucleon
down to rather small quark masses, there is no clear signal (effective mass
plateau)  for the negative parity states and for the $N_+^\prime$ at small
quark masses. However, below we will show that all states, except for the
excited positive parity state $N_+^\prime$, extrapolate  well to their physical
masses. 

For a study of excited states an analysis of finite volume effects is
particularly important. Excited states have more extended wave functions than
the ground state and thus suffer more easily from squeezing  them into a too
small volume. Such a squeezing typically increases  the kinetic energy of the
constituent quarks and so pushes up the mass of the state. In order to analyze
the finite volume effects, in Fig.\ \ref{nucleonmasses} we compare our numbers
for the baryon masses as a function of the bare quark mass $m_q$ for $16^3
\times 32$  and $12^3 \times 24$ lattices and find little volume  dependence.

Our nucleon data easily penetrate the low pion mass  region where level
crossing of the negative and positive parity states has been  claimed in
\cite{leeetal,dongetal,sasaki}.  However, we find that for the other states the
quality of our data depletes with decreasing quark masses and we could not
maintain the standards  of our fits for $a\, m_q < 0.06$. We analyzed our data
also with other methods (multi-exponential fits, Bayesian priors, maximum
entropy method) but did not obtain stable, convincing results for an excited 
positive parity state $N_+'$ below quark masses of $a\, m_q = 0.06$ as well as
for the other positive parity excited state  extracted form the $\chi_1$
correlator. Although  \cite{leeetal,dongetal,sasaki} use slightly larger
lattices (3.0\ fm as opposed to our 2.4\ fm) we cannot blame finite size
effects for the  depletion of the quality of our data, since we obtain
essentially unchanged results already on our 1.8\ fm lattice. 

The masses of the three states nucleon, $N(1535)$ and $N(1650)$ are 
extrapolated to the chiral limit. For the nucleon this is done  by a linear
extrapolation of the square of the nucleon mass $(a \, M_N)^2$ as a function of
the quark mass $a\,m_q$. For the negative parity  states we extrapolate
linearly the mass as a function of the quark mass. Since the residual 
quark mass 
is so small ($a \, m_{res} = -0.0020(5)$) it was neglected in our 
chiral extrapolation. We used fully correlated fits to take into 
account the fact that the data for different quark masses
were calculated on the same ensemble of configurations and thus are not
independent. All three fits have a
$\chi^2/d.o.f.$ less than 1.  The linear extrapolation neglects chiral
non-analyticities. We remark that we also experimented with 
functional forms suggested by quenched chiral perturbation theory \cite{LaSh}.
However, our numerical data are not sufficiently precise for a clear 
distinction 
between the simple fits described above and the form from
quenched chiral perturbation theory. Thus we decided to stay with the simpler 
2-parameter fits.

To set the scale we use either the $\rho$-mass (from
\cite{bgrlarge}) or give our results as ratios. In Table \ref{table:masses} we
list our final numbers for the masses in the chiral limit. We find that the
nucleon mass matches its physical value reasonably well 
and the two negative parity
states are about 8 \%  and 9 \% too high. Given the fact that this is a
quenched calculation the agreement of our results with experimental data is
good. In order to estimate the systematic error from the scale setting
we compare the scale from the $\rho$-mass to the scale 
from the Sommer parameter
($a_\rho = 0.154(8)$ fm, $a_{r_0} = 0.148(2)$ fm). The second error
given in the dimensionful results in the first 
line of Table \ref{table:masses}  
was computed from the difference of the two scales. 

\begin{table}[Ht]
\caption{Comparison of our $16^3 \times 32$ results in the chiral limit to 
experimental data. Masses are given in MeV, the ratios are 
dimensionless numbers.
\label{table:masses}}
\begin{ruledtabular}
\begin{tabular}{c|ccc}
 & $M_{N}$ & $M_{N(1535)}$ & $M_{N(1650)}$ 
\\ \hline 
 data   & 941(38)(28) & 1661(57)(65)     & 1796(78)(70)     \\
 exp. & 938 & 1535(20) & 1650(30) \\
\hline 
 &  & $M_{N(1535)}/M_{N}$   
 & $M_{N(1650)}/M_{N}$ \\ \hline 
 data   &     & 1.77(7) & 1.91(9) \\
 exp. &  & 1.63(2) & 1.75(3) \\
\end{tabular}
\end{ruledtabular}
\end{table}

Like other authors 
\cite{blumetal,parityminussplita,parityminussplitb,leeetal,dongetal,sasaki}. 
we also observe a smooth 
increase of the $N$ -- $N(1535)$ splitting towards the chiral limit. At large 
current quark masses the hadrons may be described as a system with orbital
motion in a color-electric confining field and the splitting is due to this
orbital excitation. The splitting slowly increases towards the chiral limit
implying the  increasing importance of another mechanism (spontaneous  chiral
symmetry breaking) in addition to confinement. This is consistent with the
chiral constituent quark model \cite{G} where an appreciable part of the
splitting is  related to the flavor-spin interaction between valence quarks.

\section{The operator content of physical states}\label{SectPhysCont}

Not only the eigenvalues of the correlation matrix $M$ in Eq.\ (\ref{matrixM})
but also the corresponding eigenvectors provide interesting information. In
particular the entries of the eigenvectors give the mixing coefficients for the
optimal operators $\widetilde{\chi}_i$ which have maximal overlap with the
physical states in terms of the original basis $\chi_i$. Let us be a little bit
more explicit. The unnormalized matrix $M_{i,j}(t)$ of Eq.\ (\ref{matrixM}) is
diagonalized by a unitary matrix $U$ which is  built from the eigenvectors
$\vec{e}^{\,(i)}$, i.e.
\begin{equation}
U \; M(t) \; U^\dagger = 
\mbox{diag} \Big( \lambda_1(t), \lambda_2(t), \lambda_3(t) \Big) \; ,
\label{diagonalization}
\end{equation}
with $U^\dagger = ( \vec{e}^{\,(1)},  \vec{e}^{\,(2)}, \vec{e}^{\,(3)} )$,
where $\vec{e}^{\,(i)}$ denotes the  $i$-th eigenvector corresponding to the
eigenvalue $\lambda_i(t)$. Inserting the expression Eq.\
(\ref{diagonalization}) into Eq.\  (\ref{matrixM}) one finds (repeated indices
are summed)
\begin{eqnarray}
\Big\langle \Big( U_{il} \, \chi_l(t) \Big) \; P^\pm \; 
\Big( U_{jk} \, \chi_k(0) \Big)^\dagger \Big\rangle & \equiv&
\langle \widetilde{\chi}_i(t) \; P^\pm \; 
\widetilde{\chi}_j(0)^\dagger \rangle \nonumber\\
 &=&  \delta_{ij} \lambda_i(t) \; .
\end{eqnarray}
Thus the optimal operators $\widetilde{\chi}_i$ which have maximal overlap
with the physical states are obtained as linear combinations
of our original operators $\chi_i$,  
\begin{equation}
\widetilde{\chi}_i = \sum_j c^{(i)}_j \, \chi_j \quad
\textrm{with} \quad  c^{(i)}_j = \vec{e}^{\,(i) \, *}_j \; .
\end{equation}
Thus the entries of the $i$-th eigenvector $\vec{e}^{\,(i)}$ provide the mixing
coefficients for the optimal operator $\widetilde{\chi}_i$  coupling to the
$i$-th state.

The coefficients $c^{(i)}_j$ are not only interesting from a technical  point
of view for constructing the optimal operator, but also provide interesting
insight into the physics of the nucleon system. These physical aspects will be
addressed in detail in the next section. 

\begin{figure}[t]
\includegraphics*[width=6.5cm]{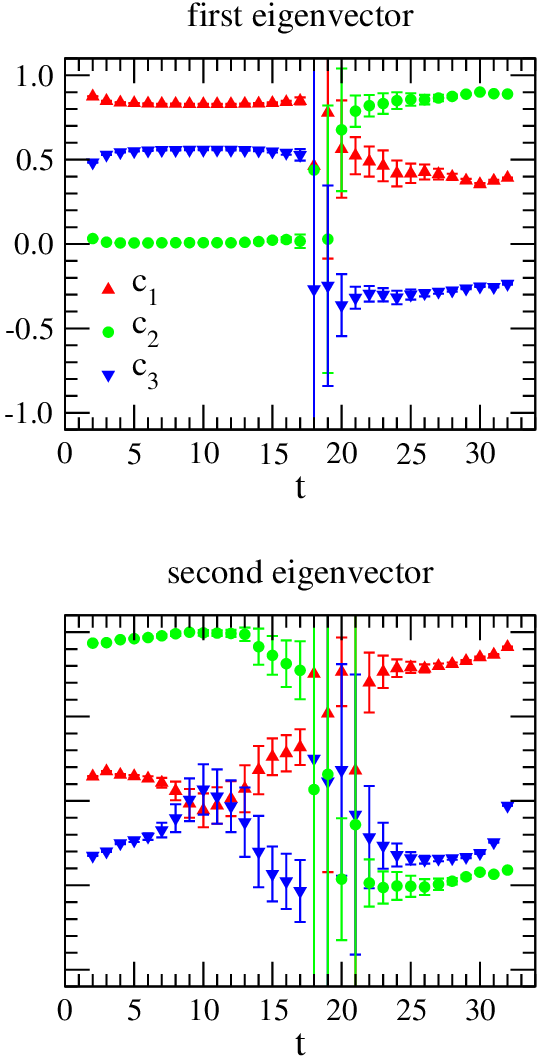}
\caption{ 
The mixing coefficients $c_i$ from the first and second eigenvector
as a function of $t$. The data are for $a\,m_q = 0.12$, $16^3 \times 32$.
\label{comps_vs_t}}
\end{figure}

With the normalization chosen for our operators $\chi_i$ the coefficients $c_j$
are real within error bars, i.e.\ the imaginary parts  are at least two orders
of magnitude smaller than the real parts and in the following we only use the
real parts \footnote{From now on it  will always be clear from the context  to
which state $i$ the coefficients $c^{(i)}_j$ correspond to and we  drop the
superscript $i$ in the following discussion.}.  In Fig.\ \ref{comps_vs_t} we
show the coefficients $c_j$ determined from the first and second eigenvector
which determine the optimal operator for the ground state and the  excited
state.

\begin{figure*}[t]
\includegraphics*[width=11.5cm]{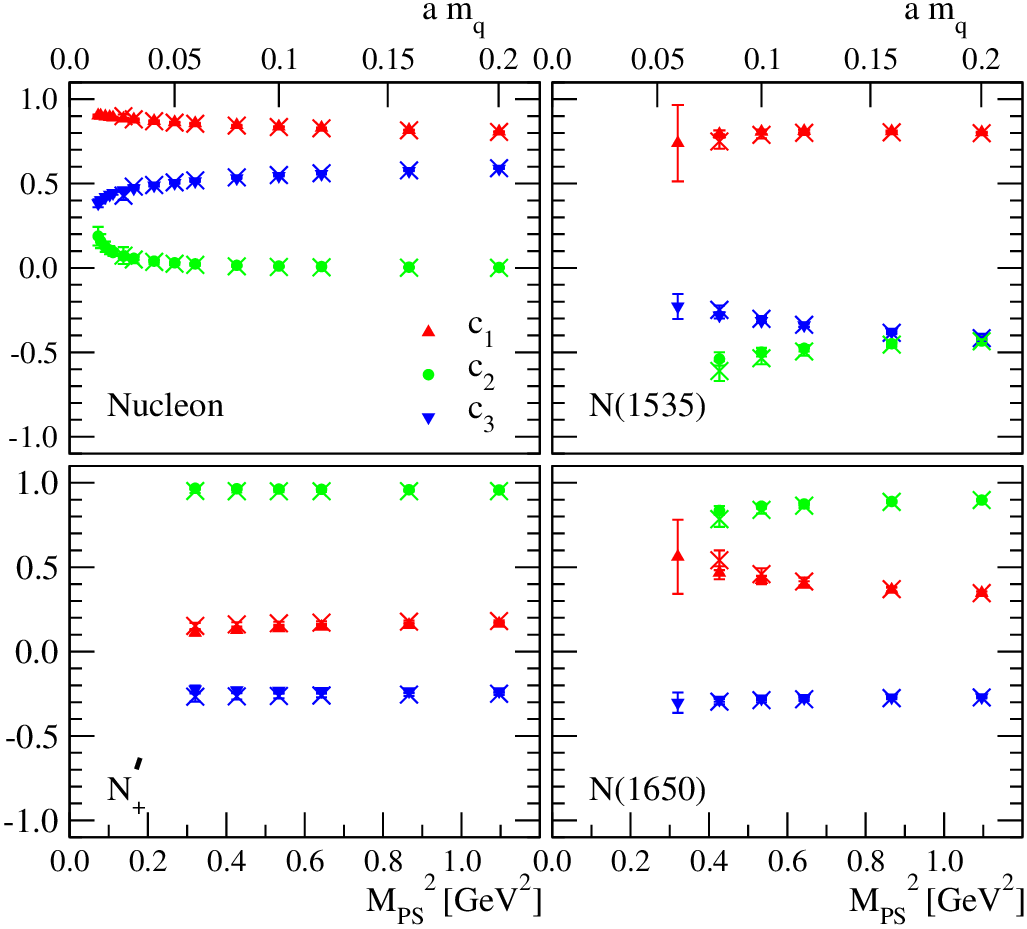}
\caption{  The mixing coefficients $c_i$ as a function of the pseudoscalar mass
squared (or dimensionless quark mass on top scale). We use large symbols for
the $16^3 \times 32$ data and superimpose the results from $12^3 \times 24$
with crosses. Note that the $c_i$ are  amplitudes and only the sum of their
squares adds up to 1, i.e.\  $|c_1|^2 + |c_2|^2 + |c_3|^2 = 1$.
\label{comps_vs_m}}
\end{figure*}

\begin{figure*}[t]
\includegraphics*[width=11.5cm]{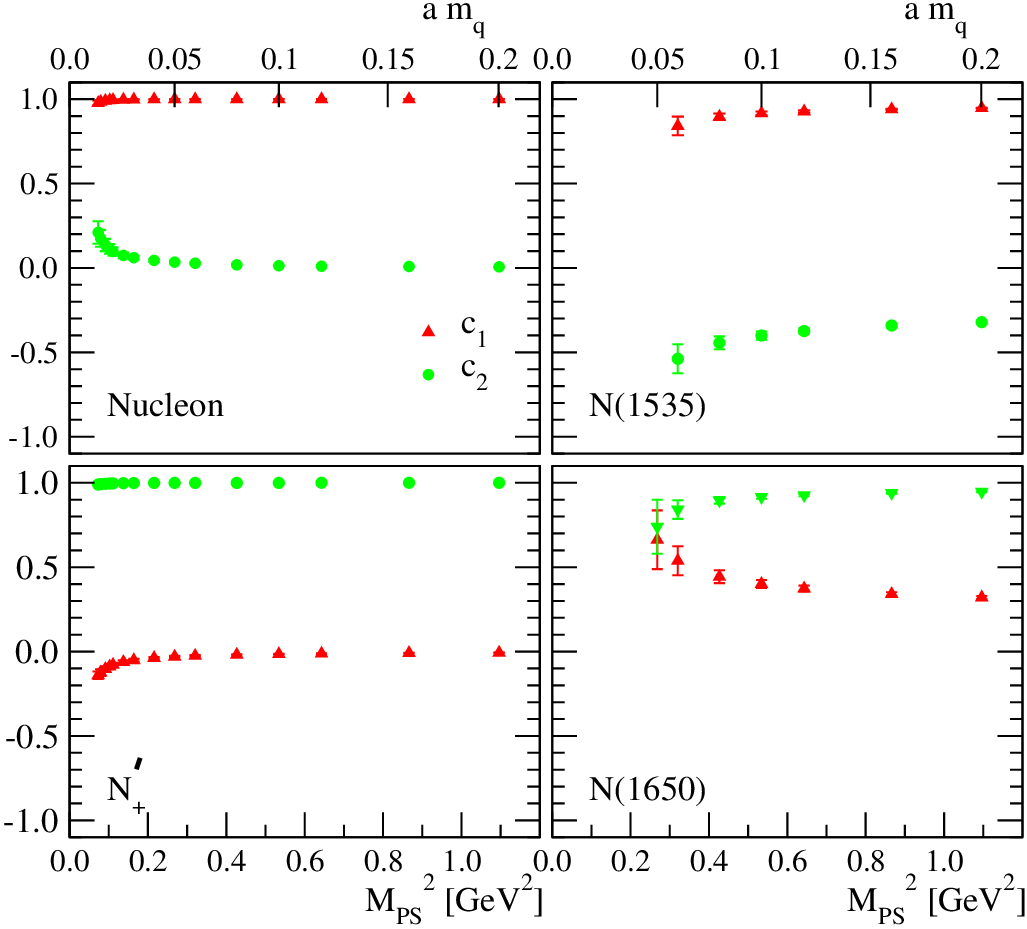}
\caption{  Like the preceding figure, but restricting the space of operators to
$\chi_1$ and $\chi_2$. The sum of the coefficient squares adds up to 1, i.e.\ 
$|c_1|^2 + |c_2|^2  = 1$.
\label{comps_vs_m2}}
\end{figure*}

Strictly speaking the matrix $U$ which is used in Eq.\ (\ref{diagonalization})
to diagonalize the correlation matrix $M(t)$ is time dependent. However,  for a
proper physical state one expects that the coefficients are  independent of $t$
within the time interval that corresponds to the given physical state.  The
time independence of the coefficients is well  pronounced for the nucleon ($t =
3 \, \ldots \, 17$ in the top plot) and also for the two negative
parity states ($t = 22 \, \ldots \, 30$).  The regions of constant $c_j$ agree
nicely with the regions where we find plateaus in the effective masses. Only in
the crossing region near $t = 19$ and close to $t=1$ for the positive parity
states, respectively $t=32$ for the negative parity  states, where higher
excited states still contribute we find strong  deviation from a constant
behavior. Only the excited positive parity state which is seen in the second
eigenvector effective mass plot of Fig.\ref{fig1} does not show long regions
with $t$-independent coefficients. It seems that this channel is for small $t$ 
dominated by an excited state between $t=2$ and $t=7$ and then changes to a
lower-lying state with different values for the $c_i$ visible until $t=17$.
This additional mixing is clearly reflected in the problems which we
encountered when trying to fit the mass of the state  $N_+^\prime$. The
corresponding low-lying artifact was discussed in Section 4 and the
corresponding structure in the correlator  was shown in Fig.\ \ref{fig2}.

An interesting question is how the coefficients depend on the  quark mass or
pion mass. In order to address this question  we take for each state at each
value of the quark mass  the values of the corresponding coefficients in the
center of the  effective mass plateau. In particular we chose $t=8$ for the
coefficients of the nucleon, $t=28$ for the two negative parity states (this
value changes to $t=20$ on the $12^3 \times 24$ lattice) and $t=5$ for the
state $N_+^\prime$. In Fig.\ \ref{comps_vs_m} we plot the coefficients as a
function of the pseudoscalar mass (the data for $M_{PS}$ were  taken from
\cite{bgrlarge}) and compare  the results from $16^3 \times 32$ (symbols) to
$12^3 \times 24$ (thin crosses)  in order to check for finite size effects. 

The absence of finite size effects which was already seen for the masses is
now   also supported by our findings for the coefficients. The results from 
$12^3 \times 24$ (crosses) agree with the data from $16^3 \times 32$ within
error bars. It is obvious that the coefficients show a non-trivial mass
dependence. The coefficients necessary for the construction of the optimal
operators $\widetilde{\chi_i}$ which couple to the $i$-th state can be read of
directly from the figure. We remark that the optimal  operators
$\widetilde{\chi_i}$ are a powerful tool which can be used to  study properties
of excited states other than just their mass such as  e.g.\ matrix elements.

\section{Physical interpretation of the mixing coefficients}
\label{SectMixingCoeffs}

As mentioned, the time component of operator $\chi_3$ couples to the spin 
$1/2$ state. In order  to check its influence we also study the ($\chi_1$,
$\chi_2$)-subsector alone. Fig. \ref{comps_vs_m2} gives the resulting mixing
coefficients analogous to Fig. \ref{comps_vs_m}. We find, that $\chi_1$ and 
$\chi_2$  behave similar in both cases and that $\chi_3$ plays a subordinate
role. Due to the different definition (e.g. our states are orthogonal and
normalized) it is difficult to compare the numbers  of Fig. \ref{comps_vs_m2} 
with the results of \cite{parityminussplita}. In both cases, as in
\cite{blumetal}, the nucleon is overwhelmingly due to $\chi_1$ while $N'_+$ is
mostly due to $\chi_2$.

First, we would like to give a simple explanation, based on a quark model picture, why
the interpolator  $\chi_2$ couples only weakly to the nucleon. According to
the quark model the nucleon belongs to the {\bf 56} representation of $SU(6)$. Its
flavor-spin wave function is completely symmetric with respect to  permutation  of two
constituent quarks. Since the color wave function is completely antisymmetric, the Pauli
principle requires that the wave function must by symmetric with respect to permutation
of the spatial coordinates of any two constituent quarks. Hence the parity of any
diquark subsystem is positive. Consequently the pure $SU(6)$ symmetric nucleon wave
function does not contain any pseudoscalar or vector diquark subsystems and the nucleon
cannot couple to the interpolator $\chi_2$. This argument is exact in the
large $N_c$ limit. 

The other question is why the nucleon couples to $\chi_1$ much stronger than
to  $\chi_3$ in spite of both having similar diquark content. The answer is
that  the interpolators belong to different chiral representations (as
discussed in  Sect.\ 3). If chiral symmetry was not broken then each hadronic
state would couple to the interpolator with concrete transformation properties
under $SU(2)_L\times SU(2)_R$. Chiral symmetry breaking implies that the
nucleon does not belong to any specific representation and couples to $\chi_1,
\chi_3$ and weakly also to  $\chi_2$, all of them belonging to distinct
representations. The stronger coupling to $\chi_1$ indicates that the
corresponding representation $(0,1/2)+(1/2,0)$ dominates the nucleon wave 
function.

The $SU(6)$ symmetry is broken by the residual spin-spin and other  interactions, which
are corrections of ${\cal O}(1/N_c)$.  The weight of the components that contain
pseudoscalar or vector diquark subsystems is  very small (e.g. for a chiral quark model
cf.\ Ref.\ \cite{GV}). This explains why  the component $c_2$ is small in the panel for
the  nucleon in Fig.\ \ref{comps_vs_m} and why one does not see the nucleon from the
interpolator $\chi_2$. While we do observe some coupling of the nucleon to the
interpolator $\chi_3$ the corresponding  amplitude $c_3$ in the upper lhs.\ plot is
small. We stress that the  $c_i$ are amplitudes, i.e.\ their squares are probabilities
adding  up to 1.

Within the quark model the Roper state belongs to the other {\bf 56} representation of
$SU(6)$. The Roper is a radial excitation  of the spatial part of the nucleon wave
function. Hence all the arguments given above for the nucleon apply also to the Roper.
However, the coefficients plot for the  $N'_+$ (lower lhs.\ plot in Fig.\
\ref{comps_vs_m}) shows a large component $c_2$ and assuming   the quark model picture
one concludes that the  $N'_+$ state seen with either the diagonal $\chi_2$-correlator
or with  the eigenvalue $\lambda_2$ is not the Roper state. A possible candidate for 
the $N'_+$ state could be some higher-lying positive parity state as e.g.\ the
$N(1710)$,  which does contain a large pseudoscalar diquark component in its wave
function.

Another important issue which we would like to discuss is the pronounced quark mass
dependence of the mixing coefficients $c_i$ for the $N(1535)$ and $N(1650)$ states,
shown in the two rhs.\ plots of Fig.\ \ref{comps_vs_m}. One clearly sees that while at
large quark masses the $N(1535)$ is dominated by $\chi_1$ and the $N(1650)$ by $\chi_2$,
the situation becomes different towards the chiral limit. This feature persists also
when studied just a subset of operators as in  Fig.\ \ref{comps_vs_m2}.  Upon
extrapolation to the chiral limit the $N(1535)$ couples optimally (up to some
uncertainty) to $(\chi_1 - \chi_2)$, while $N(1650)$ couples to $(\chi_1 + \chi_2)$ and 
the contribution of $\chi_3$ is suppressed for both negative parity states.  The very
fact that the mixing towards the chiral limit is quite different from that in the heavy
quark region implies that the physics of these resonances (and in particular of their
wave functions) is strongly influenced by chiral symmetry and its spontaneous breaking.
It is interesting to try to connect this behavior to the quark model. Both $N(1535)$ and
$N(1650)$ belong to the negative parity $L=1$ {\bf 70}-plet of $SU(6)$. Each member of
this multiplet contains in its wave function both scalar and pseudoscalar diquark
components. The mixing of  these two different components is provided by tensor forces. 

A proper mixing of two independent components does explain, in particular, peculiarities
of the $N^* \rightarrow N\eta$ decays. One can construct a linear combination such that
the $N\eta$ decay of the $N(1650)$ gets strongly suppressed, while the decay of the
$N(1535)$ gets enhanced. There are two possible origins of a tensor force providing the
necessary mixing: The tensor force which is supplied by perturbative gluon exchange  and
which does not contain flavor dependence in its operator structure \cite{ISGUR} and  the
tensor force mediated by the Goldstone field and related to spontaneous chiral symmetry
breaking \cite{G,TENSOR}\footnote{There are actually  a few different sources for this
flavor-dependent tensor force - pion-like exchange, two-pion like exchange (rho-like
exchange) etc.}. If the mixing was due to perturbative gluon exchange one would not
expect that the mixing would depend on the current quark mass. However, a tensor force
that originates from chiral symmetry breaking appears only towards the chiral limit. Our
lattice results, shown in Figs.\ \ref{comps_vs_m} and  \ref{comps_vs_m2}, clearly
support the latter possibility, though these results cannot shed any light on the
specific microscopical origin of the  tensor force. This result is also consistent with
the recent phenomenological \cite{COLL} and large $N_c$ \cite{1/N} analyses of the
mixing of the negative parity baryons: Both studies suggest that a proper mixing is
supplied by the flavor-dependent tensor force. 

Finally, our results allow to explain the smallness of the $\pi N N(1535)$ and 
$\pi N N(1650)$ coupling constants as compared to the large $\pi N N$ coupling.  It has
been shown in Ref.\ \cite{OKA} that if both the nucleon and $N(1535)$ (or $N(1650)$) are
created from the vacuum only by the $\chi_1$ and $\chi_2$ interpolators, irrespective of
their mixing, then the chiral symmetry properties of these interpolators imply that the
$\pi N N(1535)$ and  $\pi N N(1650)$ couplings must vanish. In the present work we find
that the coupling of all these baryons to the interpolator $\chi_3$, which belongs to
another representation of the chiral group, is much smaller than the coupling to
$\chi_1$ or $\chi_2$. This qualitatively explains the smallness of the $\pi N N(1535)$
and  $\pi N N(1650)$ coupling constants.

\section{Conclusions}\label{SectConcl}

Let us summarize our main findings. We use three different interpolating fields and
study how these fields couple to the nucleon and its excited states of both parities. We
clearly identify distinct negative parity excited states $N(1535)$ and $N(1650)$ and
show that the mixing of different interpolators which create these baryons from the
vacuum, is obviously quark mass dependent. The mixing towards the chiral limit is
qualitatively different from the mixing in the heavy quark region. This implies that the
physics of these states is strongly related to chiral symmetry of QCD. A very specific
mixing which is observed towards the chiral limit can be a basis for the explanation of
$N^* \rightarrow N\eta$ and $N^* \rightarrow N\pi$ decays of these baryons.

In a quark model picture we are able to understand peculiarities of the coupling of
different interpolators to the nucleon and its excited states.

We clearly see an excited state of positive parity at relatively large quark masses from
the interpolator $\chi_2$. Although we cannot reliably trace this state towards the
chiral limit we argue that it lies too  high for an identification as the Roper
resonance and may correspond to the $N(1710)$ state.

While we do observe some contribution of positive parity excited state(s) to the
$\chi_1$ interpolator at small times, the signal is rather weak and using traditional
methods, such as effective mass plots or one (two, many)-exponential fits of the
correlator, do not allow us to conclude whether this signal comes from the Roper state
or from some higher lying states. It would be an important task to optimize source and
sink in order to improve the signal from the Roper resonance.

In \cite{dongetal} the subtraction of a parametrization of possible  quenched artifacts
($\eta' N$) --- giving rise to negative values of the correlation function for central
$t$-values --- was crucial for the identification of a Roper signal. We also observe
such negative values but postpone an analysis to future studies with better statistics.

An interesting challenge for further lattice studies is the empirically observed
approximate parity doubling of nucleons above 1.7 GeV region (and similar in delta
spectrum). It has been suggested that this doubling may reflect a smooth effective
chiral symmetry restoration in the upper part of hadron spectra \cite{CG}. Still, there
remains the possibility that this doubling is accidental. Lattice methods can help to
clarify this interesting issue.

\begin{acknowledgments}
We want to thank Tom DeGrand, Meinulf G\"ockeler, Peter Hasenfratz and 
Ferenc Niedermayer for valuable discussions. The calculations were done on the Hitachi
SR8000 at the Leibniz Rechenzentrum in Munich and we thank the LRZ staff for training
and support. We acknowledge support by the Austrian Academy of Sciences (APART 654,
Ch.G.), by Fonds zur  F\"orderung der Wissenschaftlichen Forschung in \"Osterreich,
projects P14806-TPH (L.Y.G.), P16310-N08 and P16823-N08 (C.B.L. and L.Y.G.)  and by DFG
and BMBF.
\end{acknowledgments}

\end{document}